\documentclass[manuscript]{copernicus}  %82 pp a 1500
\nolinenumbers
\usepackage{color}
\usepackage[T1]{fontenc}
\usepackage{amsmath}

\begin{document}

\title{Olbert's kappa Fermi {and Bose} distribution{s}
}

{\author[1,3]{R. A. Treumann}
\author[2]{Wolfgang Baumjohann$^*$}
%\author[2]{Yasuhito Narita}
\affil[1]{International Space Science Institute, Bern, Switzerland}
\affil[2]{Space Research Institute, Austrian Academy of Sciences, Graz, Austria}
\affil[3]{Geophysics Department, Ludwig-Maximilians-University Munich, Germany\protect\\
$^*${(Correspondence to: Wolfgang.Baumjohann@oeaw.ac.at)}
}

}

\runningtitle{Olbert's kappa distributions}

\runningauthor{R. A. Treumann \& Wolfgang Baumjohann}

\received{ }
\pubdiscuss{ } %% only important for two-stage journals
\revised{ }
\accepted{ }
\published{ }

%% These dates will be inserted by the Publication Production Office during the typesetting process.

\firstpage{1}

\maketitle

%\begin{abstract}
%\noindent\textbf{Abstract}. -- 
  
%\keywords{Auroral reconnection, downward current region, auroral kilometric radiation}

%\vspace{0.5cm}
\noindent\textbf{Abstract}.-- 
%\abstract 
{The quantum version of Olbert's kappa distribution applicable to fermions is obtained. Its construction is straightforward but requires recognition of the differences in the nature of states separated by Fermi momenta. {Its complement, the bosonic version of the kappa distribution is also given, as is the procedure of how to construct a hypothetical kappa-anyon distribution. At very low temperature the degenerate kappa Fermi distribution yields a kappa-modified version of the ordinary degenerate Fermi energy and momentum. We provide the Olbert-generalized expressions of the Olbert-Fermi partition function and entropy which may serve determining all relevant statistical mechanical quantities. Possible applications are envisaged to condensed matter physics, possibly quantum plasmas,  and dense astrophysical objects like the interior state of terrestrial planets, neutron stars, magnetars where quantum effects come into play, dominate the microscopic scale but may have macroscopic consequences.} 
%\vspace{1cm}
%\end{abstract}

\section{Introduction}
The classical (non-relativistic) kappa distribution was introduced by Stan Olbert in 1966 \citep{olbert1967} and first applied in the PhD thesis of Binsack \citep{binsack1966}. \footnote{We thank C. Tsallis for kindly bringing this reference to our attention, and G. Livadiotis for the access to the thesis.}  Its first refereed  version \citep{vasyliunas1968}, applied to electron fluxes in Earth's plasma sheet (acknowledging Olbert), is usually taken as its origin \citep[For an account of its history see Chapter 1 in Ref.][]{livadiotis2017}. In the honour of Olbert we call it ``Olbert's kappa distribution'' while for simplicity shall  speak of kappa distributions below. 

The kappa distribution  occurred almost permanently when dealing with particle distributions in high temperature plasmas encountered in space \citep{christon1988,christon1989,christon1991,espinoza2018,kirpichev2020,eyelade2021}). It was also inferred from Cosmic Ray \citep{schlickeiser2002} spectra \citep[][extended into the relativistic domain]{treumann2018}, in the solar wind \citep{goldstein2005,pierrard2016,lazar2020,yoon2018,livadiotis2018}, near shocks \citep{eastwood2005,lucek2005,balogh2013}, and in the heliosphere in general \citep{fichtner2020}. Various applications to statistical probabilities in correlated systems have been reviewed as well \citep[see][for collections]{livadiotis2013,scherer2020}. It thus seems to represent a general distribution function in physical systems which in theory have been identified obeying some kind of internal correlations. Conventionally their evolution under non-stationary conditions, in particular in application to particle acceleration, is attributed to a Fokker-Planck description which naturally generates tails on the distribution function via momentum space diffusion. More basic theories have also been developed based on nonequilibrium statistical mechanics where they arise as stationary states far from thermal equilibrium \citep{treumann1999,treumann2004,treumann2008,livadiotis2018a,livadiotis2018b}. In the asymptotic limit nonlinear plasma theories \citep{hasegawa1985,yoon2005,yoon2006} provide classical distributions of this kind. To some extent kappa distributions are relatives of Tsallis' non-extensive distributions in Tsallis' thermostatistics \citep{tsallis1988} as in both formulations the entropy turns out not to be simply extensive. Tsallis' theory calls it non-extensive. Recently the analytical form of the classical (non-quantum) kappa entropy has been constructed \citep{treumann2020} which shows that it is super-extensive while being different from its thermo-statistical cousin, indicating that the fields of application in physics and statistics presumably refer to different domains. 

The appearance of non-equilibrium distributions is physically no surprise. In classical physics they arise as non-stationary distributions in {quasilinear (Fokker-Planck) theory where they can be made stationary by balancing particle injection and losses from the volume under consideration \citep{schlickeiser2002}. What however is unusual is that so far the attempts failed to construct a quantum equivalent of the classical  kappa  distribution.} Formally it seems very simple to obtained it from the partition function, but the distributions found {\citep{treumann2008,treumann2014} were rather inconvenient versions of the classical Olbertian and no real quantum distributions as they did not allow for the presence of particles with energy below Fermi energy.\footnote{{It was brought to our attention by the Associate Editor, G. Livadiotis, that in the framework of Tsallis' non-extensive statistics Fermi and Bose distributions had been derived already \citep[cf., e.g.,][]{aragao2003,christianto2007,obregon2018}. Since they arise from different propositions, they  differ from Olbert's kappa quantum distributions.}} As an explanation it was concluded that quantum theory at low temperatures suppresses correlations of the Fokker-Planck kind which cause the evolution of non-gaussian tails on the distribution. However such a conclusion is rather unsatisfactory because there is no obvious way to demonstrate that correlations are indeed excluded by quantum effects. Motivation for constructing an  kappa  quantum distribution can thus be found in the wish to correct this mismatch in order to complete and generalize the Olbertian statistical mechanical theory by extending it into the quantum domain. In the Discussion section we also provide some hints on possible applications.} An application {to non-ideal gases} which the former version was given \citep{domenech2015,domenech2020} should { for consistency probably be adapted to the correct  kappa  Fermi version of the distribution obtained below. One would expect that it improves the interpretation of the experimentally obtained results.} Here we identify the crucial step that was missed in those attempts and subsequently provide the correct  kappa  quantum distributions. We first restrict to the Fermi-Olbertian as this one is the more important in view of the wide range of quantum problems to that it may possibly be applied to some particularly suited problems.

\section{Classical  kappa  distribution}
There are several ways of deriving the classical  kappa  distribution. Consider an ideal gas of indistinguishable particles. The simplest is to start from the classical either micro-canonical or canonical Gibbs partition function  \citep{treumann2008}, {which is just the sum of all micro-canonical probabilities,} and replacing the Boltzmann factor with the Olbert factor
\begin{equation}
\exp\big(-\beta\epsilon_\alpha\big)\to \Big[1+{\beta\epsilon_\alpha/\kappa}\Big]^{-\kappa-s}
\end{equation}
Here $\epsilon_\alpha$ is particle energy in state $\alpha$, $\beta=T^{-1}$ with temperature $T$ in energy units, $0<\kappa \in \mathcal{R}$ Olbert's parameter, and $0<s$ a constant power which is eventually needed in adjusting for thermodynamical reasons. {(Note that there is no restriction on $\kappa>0$ except that it has to be positive. The analyticity of the distribution, its character of a distribution, and its correct thermodynamic properties are taken care of by $s$ in the exponent, a fixed constant number whose particular nonrelativistic and relativistic values are determined by the thermodynamic constraints  \citep[for proof see, e.g.,][]{livadiotis2013,treumann2004,treumann2014} and are given below.)} It is no problem to include a high energy cut-off $\exp(-\epsilon_\alpha/\epsilon_c)$ with $\beta\epsilon_c\gg 1$ which truncates the Olbertian at high energies. For the distribution this is not required as it converges by itself. It is however needed when calculating moments higher than the second in fluid theory \citep{scherer2017,lazar2020a,lazar2020,treumann2004}. Then on the usual way of finding the probability distribution the  kappa  distribution becomes of course trivially
\begin{equation}
p_\kappa(\epsilon_\alpha)=e^{-\epsilon_\alpha/\epsilon_c}\big[1+\frac{\beta\epsilon_\alpha}{\kappa}\big]^{-\kappa-s},\qquad \beta=T^{-1}
\end{equation}
One easily shows that this choice with $\kappa\to\infty$ reproduces Boltzmann's classical distribution where the artificial truncation is superficial and can be dropped. Summation of all occupations of states $\alpha$ yields the partition function which serves as normalization factor in the Olbert-Gibbs distribution and other purposes.  Of course, as usual, though the partition function just remains the sum of the Olbert factors, the summation still poses a major problem. 

\section{The kappa  Fermi distribution}
This same procedure fails in the quantum case in application to the Fermi or Bose distributions. One can, however, give it a different start when for simplicity referring to the slightly more lucid canonical so-called thermodynamic Gibbs potential \citep{kittel1980,landau1980} defined as the form  $d\Omega=-SdT-\mathcal{N}d\mu$ dual to energy conservation in the first law of thermodynamics {with entropy $S$, temperature $T$, particle number $\mathcal{N}$, and chemical potential $\mu$}. It is of course closely related to the canonical Gibbs partition function from which it is derived to become
\begin{equation}
\beta\Omega_\alpha=-\log\sum_{n_\alpha}\big[\exp\beta\big(\mu-\epsilon_\alpha\big)\big]^{n_\alpha}
\end{equation}
where $n_\alpha$ is the exact occupation {number} of  states $\alpha$, and the sum is to be taken over all occupations. In the Fermi case there are only two possibilities: $n_\alpha=0,1$, while the Bose case allows for arbitrary occupations. The advantage of using the thermodynamic potential is that the average occupation number simply follows from the partial derivative with respect to the chemical potential $\mu$ as
\begin{equation}
\langle n_\alpha\rangle=-\frac{\partial\Omega_\alpha}{\partial\mu} \equiv p(\epsilon_\alpha)
\end{equation}
This is the prescription for constructing the wanted distribution. However, simply substituting the Olbert factor for the Boltzmann factor in the above expression turns out to be wrong. Sometimes a very small move is necessary to progress a bit. Before proceeding, the potential must be inverted by resolving the logarithm
\begin{equation}
\exp(-\beta\,\Omega_\alpha)=\sum_{n_\alpha}\big[\exp\beta(\mu-\epsilon_\alpha)\big]^{n_\alpha}
\end{equation}
which yields a form that contains exponentials on both sides of the equation, and thus the substitution has to be done for both exponentials. Moreover, inspecting the exponential on the right one realizes that at $\mu=\epsilon_\alpha$ the character of the Boltzmann factor changes as the argument of the exponential changes sign. Physically this means that for positive chemical potential $\mu$ the exponentials of states with energy below $\mu$ behave differently from those with energy larger than $\mu$. They belong to different physical states. In Fermi theory this is automatically taken care of by the smooth behaviour of the exponential. However, in the Olbertian this smoothness is destroyed by the properties of the rational function and, as in our previous approaches \citep{treumann2004,treumann2014} who did not take this difference into account, results in non-physical distributions either forbidding occupation of any states with energy less than $\mu$  or requiring that $\mu\leq 0$ which eliminates any quantum properties. Hence, in the substitution the sign of the argument of the exponential must be adapted. This yields the expressions
\begin{equation}\label{first}
\Big[1+{\beta\Omega_\alpha/\kappa}\Big]^{-(\kappa+s)}=\left\{
\begin{array}{ccc}
 \sum_{n_\alpha}\big[1-\beta(\mu-\epsilon_\alpha)/\kappa\big]^{-n_\alpha(\kappa+s)}, & \qquad  &  \epsilon_\alpha>\mu \\[-1ex]
 && \\[-1ex]
\sum_{n_\alpha}\big[1+\beta(\mu-\epsilon_\alpha)/\kappa\big]^{-n_\alpha(\kappa+s)},   & \qquad  &   \epsilon_\alpha<\mu
\end{array} \right.
\end{equation}
which with $\mu>0$ positive, applicable to the Fermi distribution, can be written as
\begin{equation}\label{omom}
\Big[1+{\beta\Omega_\alpha/\kappa}\Big]^{-(\kappa+s)}=\sum_{n_\alpha}\big[1+\beta|\mu-\epsilon_\alpha|/\kappa\big]^{-n_\alpha(\kappa+s)}
\end{equation}
Application of the Pauli principle in order to specify to the quantum exclusions which the Fermi distribution takes care of then produces
\begin{equation}
\Big[1+{\beta\Omega_\alpha/\kappa}\Big]^{-(\kappa+s)}=1+\big[1+\beta|\mu-\epsilon_\alpha|/\kappa\big]^{-(\kappa+s)} 
\end{equation}
which must be differentiated with respect to $\mu$ keeping $\beta$ and $\epsilon_\alpha$ constant. This gives after some algebra the wanted (most probable) Olbert $\kappa$ Fermi distribution 
\begin{equation}\label{ofd-a}
\langle n_\alpha(\beta)\rangle_\mathrm{OF}=\Big\{\Big[1+|\mu-\epsilon_\alpha|\beta/\kappa\Big]^{\kappa+s}+1\Big\}^{-[1+1/(\kappa+s)]}
\end{equation}
for the average occupation number of states. It applies to both non-relativistic as well as relativistic  $\epsilon_\alpha=mc^2\gamma_\alpha$ gases with $\mathbf{p}_\alpha$ momentum and $\gamma_\alpha=\sqrt{1+p^2_\alpha/m^2c^2}$. Clearly this is not the same as if in the Fermi distribution the exponential would have been replaced with the Olbert factor. In particular the plus sign in the bracket is important as it warrants that the state of zero particle energy $\epsilon_\alpha=0$ is not forbidden at low temperature and $\mu>0$ which would be non-physical as there will always be electrons of zero energy (except in a magnetic field where the lowest state has energy $\epsilon_0=\frac{1}{2}\hbar\omega_{ce})$.  As in the ordinary Fermi distribution, the total number of particles $\mathcal{N}$ and total energy $\mathcal{E}$ are just the respective sums of all occupations and all energies
\begin{equation}
\sum_\alpha \langle n_\alpha\rangle=\mathcal{N},\qquad \sum_\alpha\epsilon_\alpha\langle  n_\alpha\rangle=\mathcal{E}
\end{equation}
For a gas of particles with momenta $\mathbf{p}$ the above discrete Olbert $\kappa$ Fermi distribution becomes a continuous function of particle energy $\epsilon_\mathbf{p}$, the Olbert $\kappa$ Fermi distribution function
\begin{equation}\label{ofd}
f_\mathrm{OF}(\mathbf{p})=\Big\{\Big[1+|\mu-\epsilon_\mathbf{p}|\beta/\kappa\Big]^{\kappa+s}+1\Big\}^{-[1+1/(\kappa+s)]}
\end{equation}
which for practical purposes when integrating over momentum and configuration space has to be properly normalized to the total number of particles  in the volume, as indicated above, using the transition from summation over discrete states to continuous integration over space and momentum $\sum_\alpha\to 2\mathcal{V}(2\pi\hbar)^{-3}\int d^3\mathbf{p}$ where one accounts for the presence of two independent electron spin directions, and the density integral is normalized to the above total particle number $\mathcal{N}$. However, while the Olbert parameter $\kappa$ accounts for deviations of the distribution, normalization by itself becomes a formidable task already for the complicated analytical form of the distribution and the subtlety of its different behaviour below and above the chemical potential $\mu$. {Applying l'H\^opital's rule it is straightforward to show that both distributions (\ref{ofd-a}) and (\ref{ofd}) for $\kappa\to\infty$ turn into their ordinary Fermi forms.}

The   kappa  Fermi distribution obtained in this way maintains the quantum properties of the gas which become most important at low temperature. For large $\kappa\gg 1$ the distribution approaches the Fermi respectively the Boltzmann distribution. At moderate $\kappa>1$ the effective temperature in the distribution is kept at a higher value thus even for low temperatures enlarging the domain of finite temperature. However, with small $\kappa<1$ this effect becomes reversed as fractional values of $\kappa$ reduce the temperature and extend the domain of low temperature into the warmer domain.  Interpreting $\kappa T$ as physical temperature would be as wrong as it was in the classical case \citep{vasyliunas1968}. It has been demonstrated elsewhere \citep[cf., e.g.,][]{yoon2005,livadiotis2013,treumann1999,treumann2004,treumann2014} that $T$ remains the real physical temperature in the classical case, a conclusion which is not violated in the quantum domain as its physical meaning is not changed by the transition from classical to quantum physics. For ideal non-relativistic and relativistic gases one in addition has $s=5/2$ and $s=4$ \citep{treumann2014,treumann2018}, respectively. 

{Asking for the shape of the  kappa  Fermi distribution at finite temperature, one excludes large $\kappa\gg1$ because it just reproduces the Fermi distribution. Medium $\kappa\sim s$ and small $\kappa\ll s$ distributions are the only interesting cases. The shape of the distribution is essentially determined by the quantities $X\equiv \beta\mu/\kappa$ and $Y=\beta\epsilon_\alpha/\kappa$ suppressing the index $\alpha$ on $Y$. One may note that for the values of interest $\kappa= z-s>0$ is positive, and $z$ is a finite number the order of a few. Moreover, the only interesting case is that of comparably large chemical potentials $\mu$ as they correspond to non-negligible Fermi energies. The limiting case $T\equiv 0$ is that of a degenerate gas which is considered below separately. In that case all states $\alpha$ below $\mu$ are occupied. For finite though low temperatures this is also the case up to a short distance $-\frac{1}{2}\Delta\epsilon$  from $\mu$. One may expand the above distribution with respect to the difference $\Delta\epsilon=|\mu-\epsilon_\alpha|$. This then yields that the distribution deviates from its value at $\epsilon_\alpha=\mu$, which is  $\langle n_\alpha(\Delta\epsilon)\rangle|_{\Delta\epsilon=0}=2^{-(1+1/z)}$, by the amount ${-\frac{1}{2}}(1+\kappa+s)\Delta\epsilon$, the energy gap, before, at larger energies $\epsilon_\alpha>\mu+\frac{1}{2}\Delta\epsilon$ changing into a power law $\epsilon_\alpha^{-(1+\kappa+s)}$ and decreasing further on.}

\begin{figure*}[t!]
\centerline{\includegraphics[width=0.75\textwidth,clip=]{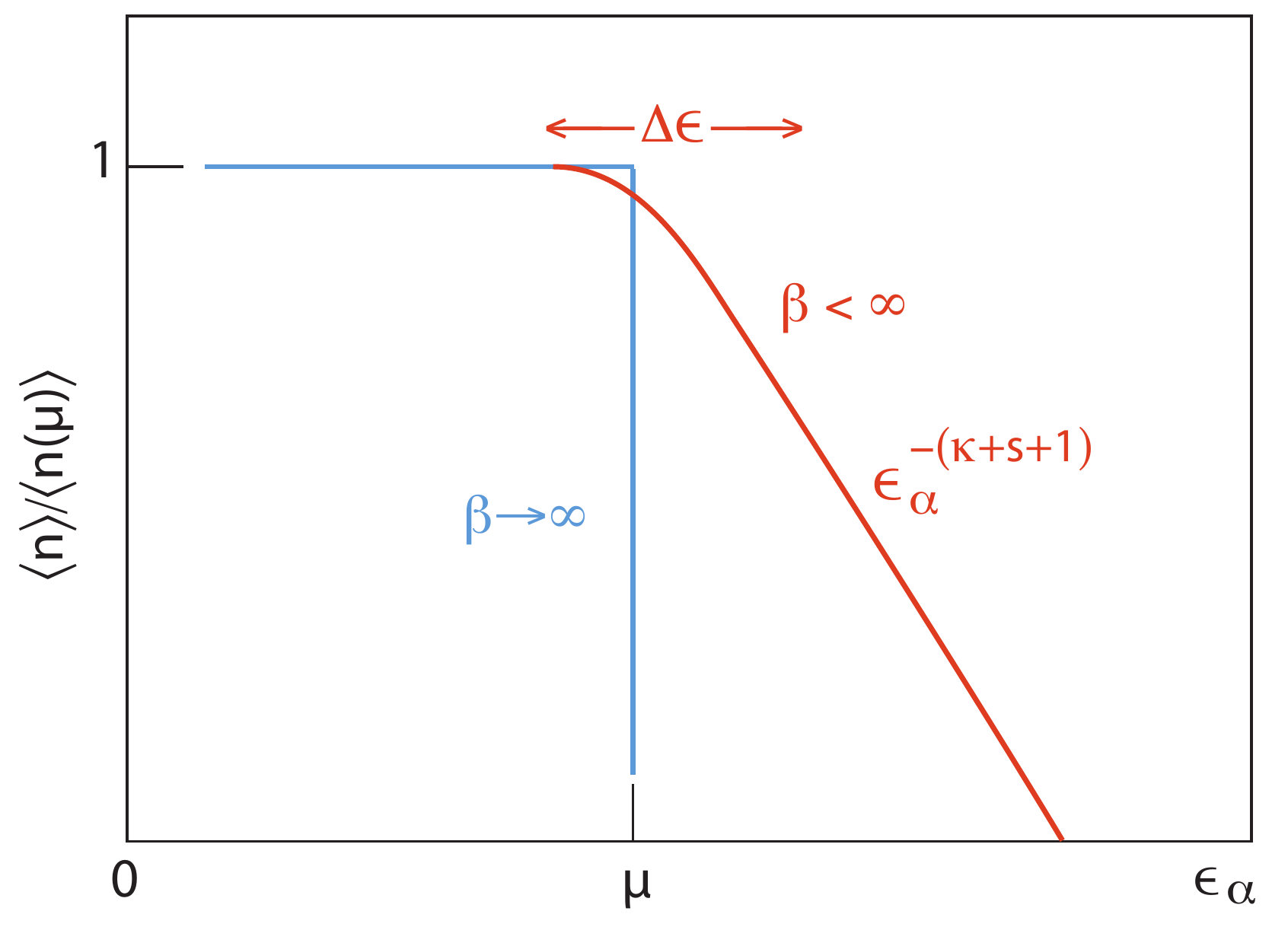}}
\caption{Schematic of  Olbert's kappa Fermi distribution as function of energy $\epsilon_\alpha$ in state $\alpha$ for the cases of zero temperature $\beta\to\infty$ (the degenerate case, shown in blue colour) and finite temperature $\beta\ll\infty$ (given in red). In the latter case of finite temperature an energy gap $\Delta\epsilon$ appears as indicated around energies $\epsilon_\alpha=\mu$ which to higher energy turns into a power law decay of the distribution. The power is a function of Olbert parameter $\kappa$ and the thermodynamic constant $s$. Note that the distribution converges for infinite energy but must be truncated in order to maintain convergence by adding an exponential cut-off factor at some energy $\epsilon_c\gg\mu$ if higher moments than mean energy $\mathcal{E}$ are required. Note the obvious similarity to the Fermi distribution which identifies  Olbert's kappa Fermi distribution as the relevant analytical generalization of the quantum mechanical Fermi distribution. The main difference is in the natural occurrence of the power law tail and the related stretching of the gap.} \label{fig1}
\end{figure*}

Figure \ref{fig1} shows a schematic of the shape of the  kappa  Fermi distribution for the two cases of zero $\beta\to\infty$ and finite $\beta\ll\infty$ temperatures.

\section{Thermodynamic potential and partition function}
In order to find a representation of the Olbert-Fermi thermodynamic Gibbs potential of an ideal gas one must invert the above distribution and insert it into the expression for $\Omega$ which yields the equivalent expressions
\begin{eqnarray}
\frac{\beta\,\Omega_\alpha^\mathrm{OF}}{\kappa}&=&\Big(1-\langle n_\alpha\rangle^\frac{\kappa+s}{\kappa+s+1}\Big)^{{1/(\kappa+s)}}-1\\
& =&\Big\{1+\big[1+\beta|\mu-\epsilon_\alpha|/\kappa\big]^{-(\kappa+s)}\Big\}^{-1/(\kappa+s)}-1 \label{omom}
\end{eqnarray}
which may serve to calculate the properties of the ideal Olbert-Fermi gas. The first part gives $\Omega_\alpha$ in terms of the average occupation number, i.e. the distribution function, the second in terms of energy $\epsilon_\alpha$. This enables to obtain the entropy which is defined as the negative derivative $S=-\partial_T\Omega$ with respect to temperature at constant volume $\mathcal{V}$ and $\mu$ or
\begin{equation}
S_\alpha=\beta^2\frac{\partial\Omega_\alpha}{\partial\beta}=\beta\Big(\frac{\partial\beta\Omega_\alpha}{\partial\beta}-\Omega_\alpha\Big)
\end{equation}
which gives  for the (dimensionless) entropy of state $\alpha$ the expression
\begin{equation}\label{entrop}
\frac{S_\alpha^\mathrm{OF}}{\kappa}=1-\sum_\pm\Big\{\Big(1+\beta|\mu-\epsilon_\alpha|/\kappa\Big)^{\pm(\kappa+s)}+1\Big\}^{-1-{1/(\kappa+s)}}
\end{equation}
One would like to have a representation of the entropy solely in terms of the mean occupation number $\langle n\rangle$. This is obtained after substantial simple though tedious algebra in the form
\begin{equation}
\frac{S_\alpha^\mathrm{OF}}{\kappa}=1-\frac{\langle n_\alpha(\beta)\rangle_\mathrm{OF}}{1-\langle n_\alpha(\beta)\rangle^{(\kappa+s)/(\kappa+s+1)}_\mathrm{OF}}
\end{equation}
which replaces the well known logarithmic form \citep{kittel1980,landau1980} of the Fermi entropy.
The total entropy $\mathcal{S}$ is obtained when summing up (\ref{entrop}) over $\alpha$ respectively integrating over momentum space.  Similarly, summing or integrating $\Omega_\alpha$ (\ref{omom})  generates the equation of state $\mathcal{PV}={-}\sum_\alpha\Omega_\alpha$ with $\mathcal{P}$ pressure, and $\mathcal{V}$ volume. Moreover, the consistent physical temperature $T\equiv\beta^{-1}$ (in energy units) is defined in the usual way as the partial derivative $\partial_\mathcal{E}\mathcal{S}=\beta$ of the total entropy with respect to the total energy $\mathcal{E}$.

{Knowing the  kappa  Fermi distribution the corresponding partition function can be written down. We here give it the canonical version which of course is identical with the above normalization condition with unspecified particle number
\begin{equation}
Z_\mathrm{OF}\equiv\sum_\alpha \langle n_\alpha\rangle_\mathrm{OF}=\sum_\alpha \Big\{\Big[1+|\mu-\epsilon_\alpha|\beta/\kappa\Big]^{\kappa+s}+1\Big\}^{-[1+1/(\kappa+s)]}
\end{equation}
As usual, it can be used to evaluate all thermodynamically interesting quantities.}

{\section{The kappa  Bose distribution}
For completeness, we note in passing that the same procedure allows obtaining the complementary Olbert-Bose potential summing up the right-hand side of  Eq. (\ref{first}) for $0\leq n_\alpha<\infty$ . Summation yields
\begin{equation}\label{obpot}
\beta\Omega_\alpha^\mathrm{OB}/\kappa=\Big\{1-\big[1-\beta(\mu-\epsilon_\alpha)/\kappa\big]^{-(\kappa+s)}\Big\}^{1/(\kappa+s)}-1
\end{equation}
which has to be differentiated with respect to $\mu$ holding $\epsilon_\alpha$ constant in order to obtain the average occupation number $\langle n_\alpha\rangle_\mathrm{OB}$ of states of an Olbert-Bose gas. This gives immediately 
\begin{equation}
\langle n_\alpha\rangle_\mathrm{OB}=\Big\{\big[1-\beta(\mu-\epsilon_\alpha)/\kappa\big]^{(\kappa+s)}-1\Big\}^{-1+1/(\kappa+s)}\big[1-\beta(\mu-\epsilon_\alpha)/\kappa\big]^{-2}
\end{equation}
which must be normalized to the total particle number. Clearly, for finite temperature $\beta\ll\infty$ the chemical potential $\mu\leq0$ must be negative as follows from the requirement that the occupation of the ground state $\epsilon_\alpha=0$ is not forbidden. Like in the Fermi case its entropy and equation of state can be obtained straightforwardly. We do not pursue this case here in any detail. We just point out that in contrast to ordinary quantum statistics, the  kappa  Fermi and -Bose distributions are not  mirror symmetrical versions of each other. }

The kappa Bose distribution possesses a finite occupation number at zero energy $\epsilon_\alpha=0$ given by
\begin{equation}
\langle n_\alpha(0)\rangle_\mathrm{OB}=\Big\{\big[1+\beta|\mu|/\kappa\big]^{(\kappa+s)}-1\Big\}^{-1+1/(\kappa+s)}\big[1+\beta|\mu|/\kappa\big]^{-2}
\end{equation}
with $\mu<0$ and $\beta$ finite which is the same for all states $\alpha$. Since all states contribute to it, summing over states implies that the state $\epsilon_\alpha=0$ is condensed. The kappa Bose distribution thus also exhibits Bose-Einstein condensation. Since the power $\kappa+s$ is not an integer, the first factor indicates that, at finite particle number, reality and finiteness of the condensate require vanishing of the chemical potential $|\mu|\to 0$ for $\beta\to\infty$ at zero temperature. With $\beta|\mu|$ finite, the second factor then causes a reduction of the condensate. 

Investigation of other effects caused by the Olbert kappa Bose distribution will be of interest. For instance, applying it to massless bosons like photons can be done easily as for them $\mu=0$ and energy $\epsilon_\alpha=\hbar\omega_\alpha$ is replaced in the distribution. This enables calculating the mean energy by integrating over the frequency, and other quantities in photon gases.

{Again, the Olbert-Bose partition function is the sum of all the un-normalized Olbert $\kappa$ Bose distributions in states $\alpha$
\begin{equation}
Z_\mathrm{OB}\equiv\sum_\alpha\langle n_\alpha\rangle_\mathrm{OB}=\sum_\alpha\frac{\Big\{\big[1-\beta(\mu-\epsilon_\alpha)/\kappa\big]^{(\kappa+s)}-1\Big\}^{-1+1/(\kappa+s)}}{\big[1-\beta(\mu-\epsilon_\alpha)/\kappa\big]^{2}}
\end{equation}}

{One may, in addition, note that the thermodynamic Gibbs potential is an energy. It thus remains additive (extensive) independent of any other properties of the gas. Hence on the way of fractionally adding the Olbert Fermi and Bose potentials $\Omega_\alpha=q\Omega_\alpha^\mathrm{OF}+(1-q)\Omega_\alpha^\mathrm{OB}$ with $q<1$ and differentiating with respect to $\mu$ and $\beta$ one also can construct the mixed Olbert $\kappa$ anyon distribution and entropy of Olbert anyon gases, applicable  to the fractional quantum-Hall effect, for instance, where correlations are known to exist even on the quantum level.}

\section{Degenerate Olbert-Fermi gas}
The most interesting is the degenerate electron gas with temperature $T\approx0$, more generally with $T\ll\mathcal{E}_\mathrm{OF}$, where $\mathcal{E}_\mathrm{OF}$ is the Olbert-Fermi energy defined below.  With $\kappa\to\infty$ the Eqs. (\ref{ofd-a}) and (\ref{ofd}) are of course the usual degenerate Fermi distributions. Leaving $\kappa\ll\infty$ and $T\approx0$ Eqs. (\ref{ofd-a},\ref{ofd}) become the degenerate Olbert $\kappa$ Fermi distributions. Like in the Fermi distribution, for $T\to0$ all energy levels $\alpha$ become confined below $\mu$, the Fermi energy. Here this gives for the total particle density, when interpreting the sum as an integral with respect to the particle momentum $p$ and integrating the  kappa  Fermi distribution (\ref{ofd}) up to $\mu=p_\mathrm{OF}^2/2m$ counting two spins per state,
\begin{equation}\label{eq-N}
N=\frac{1}{\pi^2\hbar^3A_\kappa}\int\limits_0^{p_\mathrm{OF}}p^2dp=\frac{p_\mathrm{OF}^3}{3A_\kappa\pi^2\hbar^3},\qquad A_\kappa=2^{1/(\kappa+s)}
\end{equation}
This yields the expressions for $p_\mathrm{OF}$, the Olbert-Fermi moment, and $\epsilon_\mathrm{OF}$, the Olbert-Fermi energy
\begin{equation}
p_\mathrm{OF}=A_\kappa^\frac{1}{3} p_F,\qquad \epsilon_\mathrm{OF}=\frac{p_F^2}{2m}A_\kappa^\frac{2}{3}, \quad\mathrm{where}\quad p_F=\hbar k_F
\end{equation}
and $k_F=(3\pi^2 N)^{1/3}$ is the Fermi wavenumber.  As one observes, the effects of the Olbertian transformation on the properties of the degenerate Fermi gas are moderate. They become susceptible only at small $\kappa<1$ and at those numbers are determined mainly by the value of $s=5/2$ thus being practically constant. Following Eq. (\ref{eq-N}),  when $\kappa$ approaches zero and $s=5/2$, then $A_\kappa$ approaches 1.36 (roughly 1.4). Therefore the effect of $\kappa$ is to stretch or shrink the quantum domain of the ideal gas. For large $\kappa\gg1$ there is no change as this becomes the Fermi case. These expressions can be used to calculate 
\begin{equation}
\mathcal{E}_\mathrm{OF}=\frac{A_\kappa^{2/3}p_\mathrm{F}^{5/3}}{5\pi^2\hbar^3}, \qquad\mathrm{and}\qquad \mathcal{P}_\mathrm{OF}=\frac{2}{15}\frac{A_\kappa^{2/3}p_\mathrm{F}^{5/3}}{\pi^2\hbar^2}
\end{equation}
the mean energy $\mathcal{E}$ and degeneracy pressure $\mathcal{P}$ of the electron gas in the usual way,  which turn out to become simple $A_\kappa$-modifications of the usual properties of a degenerate Olbert-Fermi gas. At small but finite temperatures the distribution bents down from its constant value below $\epsilon_\alpha=\mu$ at distance $-\frac{1}{2}\Delta\epsilon$ from the Fermi boundary to cross the Fermi boundary until at $+\frac{1}{2}\Delta\epsilon$ becoming power law $\propto\epsilon_\alpha^{-(1+\kappa+s)}$ and decaying to low values. This is similar to the ordinary Fermi case though without the common exponential decay of the distribution. The latter is replaced by the power law decay causing an interesting substantial increase of the gap. One may note that the distribution converges at large $\epsilon_\alpha$. However when moments larger than the mean energy are required, then the distribution has to be truncated through introducing an exponential high-energy cut-off $\exp(-\epsilon_\alpha/\epsilon_c)$ with $\epsilon_c\gg\mu$, as indicated in the section on the classical Olbertian. On the other hand, the distribution function of the Olbert-Fermi gas can be used in the calculation of the modified thermal and magnetic properties of the Olbert-Fermi gas and in any other application like degenerate stars.
  
\section{Summary}
{In this Brief Communication we generalized, based on physical arguments, the classical  kappa  distribution to fermions and bosons, the  quantum Fermi and Bose cases.  We also noted, how this can be extended to mixed fermion-boson anyon states. The distributions, Gibbs potentials, and partition functions can be used to derive all relevant thermodynamic and statistical mechanical quantities for quantum and classical systems. In principle this completes and rounds up the task of a generalized Olbertian statistical mechanics which turns out in all cases to be the analytical generalization of Boltzmann-Gibbs statistical mechanics into the Lorentzian and quantum domains. It results in well defined distributions, Gibbs potentials, partition functions, and entropies. If required, it can under appropriate conditions serve as the thermodynamic theory for both classical and quantum systems. As such it does apply to any system, not being restricted to plasmas. In fact, plasmas may not be the most interesting of such systems. It seems that the main field of application could be envisaged in some kinds of solid state physics, condensed matter physics, suitable astrophysical object, and possibly also, if relevant, high energy physics. It should be noted that even under quantum conditions the thermodynamic definition of temperature is maintained, which is a satisfactory physical fact.} 

For finite temperatures, the  kappa  Fermi distribution yields a substantial modification which, however, in the degenerate case of near-zero temperature, i.e. environmental temperature  much below Fermi temperature $\beta\mathcal{E}_\mathrm{OF}\gg1$, which in solid state physics and astrophysical applications is the most interesting case, just leads to some more or less slight modifications of the well-known expressions of Fermi energy and momentum. The Olbert parameter $\kappa$ accounts for deviations of the distribution from fermionic  standard indicating the presence of either internal correlations or degrees of freedom if applicable. These affect the low temperature behaviour though, as it seems, only weakly and,  preferably for small  $\kappa$, increase the quantum domain a bit into the higher energy/temperature range by shifting the Fermi energy. Investigation of the higher temperature properties is more complicated by the analytical form of the  kappa  Fermi distribution. This has to be done case by case if required as it might have some effect for instance in the generation of the gap in superconductivity theory. 

{We should, however, note at this occasion that the above investigation of the finite temperature behaviour of the  kappa  Fermi distribution with increase of the gap around Fermi energy might be of interest in superconductivity theory. It still awaits an extension of BCS-theory \citep{bardeen1957} to the application of the Olbert-Fermi statistical mechanics.}

{For completeness the complementary Olbert-Bose thermodynamic potential and distribution have also been given. The thermodynamic potential may serve to obtain the Olbert-Bose  entropy, and equation of state. In this case there is no indication of any presence of a Bose-Einstein condensation such that one would conclude that internal quantum correlations hidden in a finite value of $\kappa$ destroy the formation of a bosonic quantum condensate at zero physical temperature.}

{Classical Olbertian statistical mechanics, as by now known (for reference the reader is referred to the literature cited in the introduction) applies to high temperature collisionless plasmas as encountered in dilute near-Earth space, the heliosphere, and probably also environments of  stars and planetary systems.  kappa  distributions of particles have in the past  been inferred in a multitude of space plasmas. In dilute high temperature plasma, similar to ordinary quantum physics, quantum effects are wiped out. Application of Fermi and Bose Olbertians and the corresponding Olbertian physics there becomes clearly obsolete. Quantum plasmas, on the other hand, may provide a promising field of application but require the transition from the currently practized fluid approach based on Madelung-Bohm theory to an appropriate statistical mechanics. Quantum plasmas are dense enough in this case, have very high Fermi temperature which by far  exceeds the physical temperature $\beta^{-1}$. Physical examples are found in compact astrophysical objects like neutron stars, pulsars, magnetars, active galactic nuclei, and also black holes (if anything can be inferred about their interior state other than collapse). Electrons in the overlapping energy bands of the compact objects become, say, asymptotically free in the sense that they can freely move, resembling an electron fluid which may be subject to Olbert-Fermi statistics.}

{In similar sense a system closer-by and of large interest are terrestrial planetary interiors, for instance Earth's outer core. Its temperature is some 3500 K corresponding to an energy of $\beta^{-1}\sim 0.3$ eV and Fermi energy of conduction electrons (assuming $^{56}$Fe ions as basic constituent) of $\epsilon_F> $ few eV. Hence either Fermi or Olbert-Fermi physics applies. It may have consequences in planetary dynamo theory taking into account statistical microphysics in the excitation of the magnetic field, and if generating superconductivity, may presumably lead to the generation of a Ginzburg-Landau-Meissner effect of chains of magnetic bubbles/depletions in the magnetized outer core medium, an effect that should affect the interior structure of planetary magnetic fields. } 

{This research was driven by curiosity, not by any applicational needs. It just presents a generalization of known classical statistical mechanics into the quantum domain. It may await application to the above indicated fields.}

\end{document}